\documentclass[referee,pdflatex,sn-nature]{sn-jnl}% referee option is meant for double line spacing

\usepackage[mathlines]{lineno}
\usepackage{graphicx}%
\usepackage{multirow}%
\usepackage{amsmath,amssymb,amsfonts}%
\usepackage{amsthm}%
\usepackage{mathrsfs}%
\usepackage[title]{appendix}%
\usepackage{xcolor}%
\usepackage{textcomp}%
\usepackage{manyfoot}%
\usepackage{booktabs}%
\usepackage{algorithm}%
\usepackage{algorithmicx}%
\usepackage{algpseudocode}%
\usepackage{listings}%
\usepackage{bookmark}%
\usepackage{siunitx}
\DeclareSIUnit\pixel{px}
\usepackage{acro}
\DeclareAcronym{MBI}{
	short=MBI,
	long=Max Born Institute
}

\DeclareAcronym{CCD}{
	short=CCD,
	long=charge-coupled device
}

\DeclareAcronym{CMOS}{
	short=CMOS,
	long=complementary metal-oxide semiconductor
}

\DeclareAcronym{MOENCH}{
	short=MOENCH,
	long=Micropixel with enhanced pOsition rEsolution usiNg CHarge integration
}

\DeclareAcronym{SNR}{
	short=SNR,
	long=signal-to-noise ratio,
}

\DeclareAcronym{GMR}{
	short=GMR,
	long=giant magneto resistance,
}

\DeclareAcronym{BBO}{
	short=BBO,
	long=Beta Barium Borat (BaB$_2$O$_4$),
}

\DeclareAcronym{MLB}{
	short=MLB,
	long=magnetic linear birefringence,
}

\DeclareAcronym{MLD}{
	short=MLD,
	long=magnetic linear dichroism,
}

\DeclareAcronym{MCB}{
	short=MCB,
	long=magnetic circular birefringence,
}

\DeclareAcronym{MCD}{
	short=MCD,
	long=magnetic circular dichroism,
}

\DeclareAcronym{XMLB}{
	short=XMLB,
	long=X-ray magnetic linear birefringence,
}

\DeclareAcronym{XMLD}{
	short=XMLD,
	long=X-ray magnetic linear dichroism,
}

\DeclareAcronym{XMCB}{
	short=XMCB,
	long=X-ray magnetic circular birefringence,
}

\DeclareAcronym{XMCD}{
	short=XMCD,
	long=X-ray magnetic circular dichroism,
}

\DeclareAcronym{RMXS}{
	short=RMXS,
	long=resonant magnetic \ac{sxray} scattering
}

\DeclareAcronym{SAXS}{
	short=SAXS,
	long=small-angle X-ray scattering
}

\DeclareAcronym{RKKY}{
	short=RKKY,
	long=Ruderman–Kittel–Kasuya–Yosida,
}

\DeclareAcronym{DMI}{
	short=DMI,
	long=Dzyaloshinskii–Moriya interaction,
}

\DeclareAcronym{FEL}{
	short=FEL,
	long=free-electron laser,
}

\DeclareAcronym{XFEL}{
	short=XFEL,
	long=X-ray free-electron laser,
}

\DeclareAcronym{SOC}{
	short=SOC,
	long=spin orbit coupling,
}

\DeclareAcronym{EUV}{
	short=EUV,
	long=extreme ultra violet,
}
\DeclareAcronym{FM}{
	short=FM,
	long=ferromagnet,
}
\DeclareAcronym{AFM}{
	short=AFM,
	long=anti-ferromagnet,
}
\DeclareAcronym{TM}{
	short=TM,
	long=transition metal,
}
\DeclareAcronym{RE}{
	short=RE,
	long=rare earth,
}
\DeclareAcronym{XLD}{
	short=XLD,
	long=X-ray linear dichroism,
}

\DeclareAcronym{XAS}{
	short=XAS,
	long=X-ray absorption spectroscopy,
	long-plural-form=X-ray absorption spectroscopies,
}

\DeclareAcronym{MOKE}{
	short=MOKE,
	long=magneto optical Kerr effect,
}

\DeclareAcronym{TMOKE}{
	short=T-MOKE,
	long=transversal \ac{MOKE},
}

\DeclareAcronym{CPL}{
	short=CPL,
	long=circularly polarised light,
}
\DeclareAcronym{LPL}{
	short=LPL,
	long=linearly polarised light,
}
\DeclareAcronym{RPL}{
	short=RPL,
	long=randomly polarised light,
}

\DeclareAcronym{EPL}{
	short=EPL,
	long=elliptically polarised light,
}

\DeclareAcronym{LSF}{
	short=LSF,
	long=large-scale facility,
	long-plural-form = large-scale facilities
}
\DeclareAcronym{HHG}{
	short=HHG,
	long=higher harmonic generation,
}

\DeclareAcronym{LPXS}{
	short=LPXS,
	long=laser-driven plasma X-ray source
}

\DeclareAcronym{PXS}{
	short=PXS,
	long=plasma X-ray source
}

\DeclareAcronym{sxray}{
	short=sX-ray,
	long=soft-X-ray
}

\DeclareAcronym{XUV}{
	short=XUV,
	long=extreme ultraviolet
}

\DeclareAcronym{IR}{
	short=IR,
	long=infrared
}
\DeclareAcronym{ML}{
	short=ML,
	long=multilayer
}

\DeclareAcronym{SL}{
	short=SL,
	long=super lattice
}

\DeclareAcronym{NEXAFS}{
	short=NEXAFS,
	long=near edge X-ray absorption and fine structure
}

\DeclareAcronym{DOS}{
	short=DOS,
	long=density of states
}

\DeclareAcronym{TF}{
	short=TF,
	long=thin film
}

\DeclareAcronym{IP}{
	short=IP,
	long=in plane
}

\DeclareAcronym{OOP}{
	short=OOP,
	long=out of plane
}

\DeclareAcronym{ns}{
	short=ns,
	long=nanosecond,
}

\DeclareAcronym{ps}{
	short=ps,
	long=picosecond,
}

\DeclareAcronym{fs}{
	short=fs,
	long=femtosecond,
}

\DeclareAcronym{RZP}{
	short=RZP,
	long=reflective zone plate,
}

\DeclareAcronym{SR}{
	short=SR,
	long=synchrotron radiation,
}

\DeclareAcronym{MFM}{
	short=MFM,
	long=magnetic force microscopy,
}

\usepackage{lmodern}
\unnumbered% uncomment this for unnumbered level heads

\newcommand{\qfirst}{q_\mathrm{1st}}

\begin{document}

	\title{Laser-driven resonant soft-X-ray scattering for probing picosecond dynamics of nanometre-scale order}
	
	\author[1]{\fnm{Leonid} \sur{Lunin}}\equalcont{These authors contributed equally to this work}
	\author[1]{\fnm{Martin} \sur{Borchert}}\equalcont{These authors contributed equally to this work}
	\author[1]{\fnm{Niklas} \sur{Schneider}}
	\author[1]{\fnm{Konstanze} \sur{Korell}}
	\author[1]{\fnm{Michael} \sur{Schneider}}
	\author[1]{\fnm{Dieter} \sur{Engel}}
	\author[1,2]{\fnm{Stefan} \sur{Eisebitt}}
	\author*[1]{\fnm{Bastian} \sur{Pfau}}\email{pfau@mbi-berlin.de}
	\author*[1]{\fnm{Daniel} \sur{Schick}}\email{schick@mbi-berlin.de}
	
	\affil[1]{\orgname{Max-Born-Institut f{\"u}r Nichtlineare Optik und Kurzzeitspektroskopie}, \orgaddress{\street{Max-Born-Stra{\ss}e 2A}, \postcode{12489} \city{Berlin}, \country{Germany}}}
	\affil[2]{\orgname{Technische Universit{\"a}t Berlin}, \orgdiv{Institut f{\"u}r Optik und Atomare Physik}, \orgaddress{\street{Stra{\ss}e des 17.\ Juni 135}, \postcode{10623} \city{Berlin}, \country{Germany}}}

	\date{\today}
	
	\abstract{ % 200 words, unreferenced
		X-ray scattering has been an indispensable tool in advancing our understanding of matter, from the first evidence of the crystal lattice to recent discoveries of nuclei's fastest dynamics.
		In addition to the lattice, ultrafast resonant elastic scattering of soft X-rays provides a sensitive probe of charge, spin, and orbital order with unparalleled nanometre spatial and femto- to picosecond temporal resolution.
		However, the full potential of this technique remains largely unexploited due to its high demand on the X-ray source.
		Only a selected number of instruments at large-scale facilities can deliver the required short-pulsed and wavelength-tunable radiation, rendering laboratory-scale experiments elusive so far.
		Here, we demonstrate time-resolved X-ray scattering with spectroscopic contrast at a laboratory-based instrument using the soft-X-ray radiation emitted from a laser-driven plasma source.
		Specifically, we investigate the photo-induced response of magnetic domains emerging in a ferrimagnetic FeGd heterostructure with \SI{9}{ps} temporal resolution.
		The achieved sensitivity allows for tracking the reorganisation of the domain network on pico- to nanosecond time scales in great detail.
		This instrumental development and experimental demonstration break new ground for studying material dynamics in a wide range of laterally ordered systems in a flexible laboratory environment.
	}
	
	
	\maketitle
	
	Phases of laterally extended nanometre-scale textures of charge, orbital, and magnetic order play an important role in the physics of complex materials with competing interactions~\cite{Das2019, da_silva_neto2014, Hellwig2007, Dagotto2005}.
	Understanding the structure and dynamics of these phases requires tools that provide resolution on the relevant length and time scale combined with the flexibility to study the materials in tailored environments while applying strong and ultrafast excitations.
	Soft X-rays are perfectly suited to investigate these ordering phenomena, as their penetration depth and sensitivity are sufficiently high to probe through the entire thickness of a nanoscale material and generate contrast in ultrathin films.
	
	Emergent textures are often incommensurate with the crystal lattice and usually lack long-range order.
	Therefore, X-ray scattering from these textures typically leads to a diffuse signal, commonly referred to as \ac{SAXS}, as it appears at smaller scattering vectors than crystallographic Bragg scattering.
	In particular, when tuning the X-ray wavelength in resonance with electronic transitions of the material's constituents, \ac{SAXS} provides unique opportunities to study the nanoscale structure and dynamics of emergent phases with spectroscopic selectivity.
	Magnetic contrast for X-ray scattering is provided by tuning the photon energy close to the vicinity of spin-orbit-split core-to-valence-band transitions.
	These resonances are especially pronounced in the soft-X-ray range, for photon energies matching the L$_{2,3} (2p_{1/2}$ or $2p_{3/2} \rightarrow 3d$) absorption edges of the \acp{TM}, as well as the M$_{4,5} (3d_{3/2}$ or $3d_{5/2} \rightarrow 4f$) edges of the \acp{RE}, corresponding to \SIrange{1}{2}{nm} X-ray wavelength~\cite{Kortright2001}. 
	
	A resonant soft-X-ray scattering experiment has high demands on the photon source used.
	The wavelength must be tunable in a wide range, going substantially higher than \SI{1}{keV} to cover the magnetically relevant \ac{TM} and \ac{RE} elements.
	%The detection space is inherently three-dimensional (a 2D reciprocal space and the time domain). 
	The scattering signal drops dramatically for high spatial resolution corresponding to large scattering vectors, $q$, with $q^{-3}$ to $q^{-4}$, and lateral contrast variations in the material are often small, particularly in ultrafast time-resolved measurements.
	Consequently, their time-resolved detection requires high photon numbers compressed into pulses of ultra-short duration.
	So far, such experiments have exclusively been possible at \acp{XFEL} when they aim at matching the intrinsic timescales of the emerging order of femto- to picoseconds.
	For a long time, the technical demands were beyond the reach of laboratory instruments, which would overcome the limitations in accessibility and flexibility of \ac{XFEL} installations.
	
	We have recently developed a laser-driven \ac{PXS} operating in the relevant spectral range at a temporal resolution of $\SI{9}{ps}$~\cite{Borchert2023b}.
	The performance of the source has already been demonstrated for resonant soft-X-ray diffraction in reflection~\cite{Schick2021b} and spectroscopic \ac{XMCD} experiments in transmission~\cite{Borchert2023}.
	Based on this source, we have now developed the instrumentation to detect even weaker diffuse resonant scattering, which was, so far, inaccessible in a laboratory setup for photon energies across the \ac{TM} L and \ac{RE} M edges.
	Here, we demonstrate magnetic \ac{SAXS} at this laser-driven source up to photon energies above \SI{1}{keV} and benchmark the capabilities of our instrumentation in a full-scale time-resolved \ac{SAXS} experiment.
	
	\section{Results}
	
	As a realistic test case for all relevant instrumental parameters, we study the light-induced magnetisation dynamics in a multilayer of the \ac{TM} Fe and the \ac{RE} Gd.
	The alternating ferromagnetic Fe and Gd layers couple antiferromagnetically to each other. 
	Due to the competition of magnetic interactions in this material system, low-energetic external stimuli, e.g., magnetic fields and temperature, can lead to the formation of complex spin structures such as nanoscale magnetic domains~\cite{Hintermayr2021} or twisted spin states~\cite{Drovosekov2019}.
	Furthermore, ferrimagnetic \ac{RE}--\ac{TM} systems are technologically attractive due to their ability for all-optical switching \cite{Radu2011, Davies2020} and to host highly mobile and ultrasmall spin textures~\cite{Montoya2017, Titze2024}. 
	Our multilayer exhibits magnetic stripe domains at room temperature, forming a maze pattern with alternating up and down out-of-plane magnetisation with a spatial period of $d \approx \qty{465}{nm}$; see \ac{MFM} image in Fig.~\ref{fig:setup}b.
	
	\subsection{The laboratory soft-X-ray scattering instrument}
	
	\begin{figure*}[htbp]
		\centering
		\includegraphics[width=1.0\linewidth]{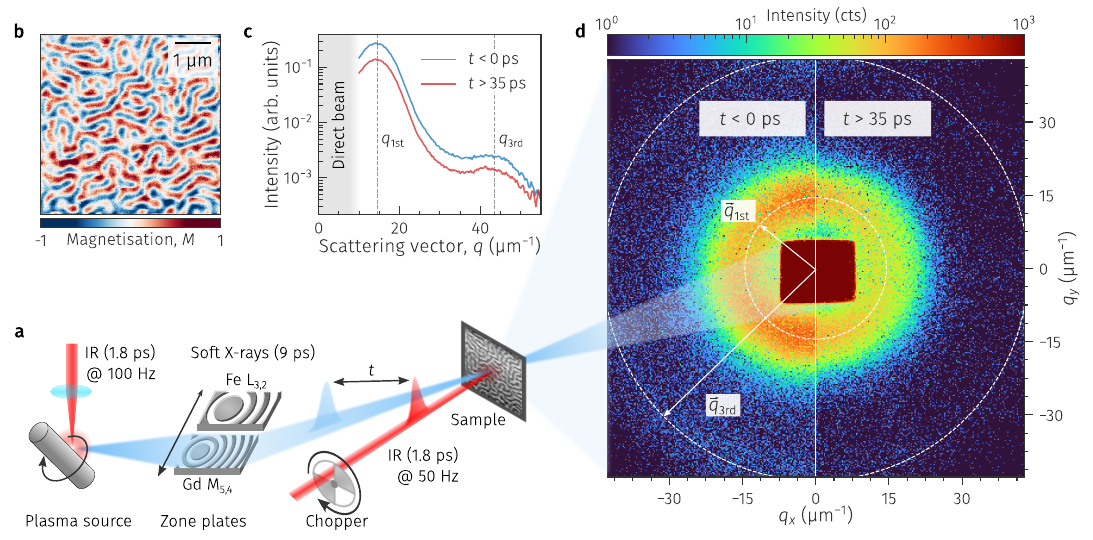}
		\caption{\textbf{Laboratory-based time-resolved SAXS experiment.}
			\textbf{a},~Sketch of the laser-driven plasma source and the scattering setup.
			The \ac{SAXS} pattern detected by a fast-readout area detector together with the direct beam is shown in panel \textbf{d}.
			\textbf{b},~Magnetic force microscopy image of the magnetic domain pattern in the FeGd multilayer.
			\textbf{c,d},~The time-resolved \ac{SAXS} signal detected at the Gd~M$_5$ absorption edge (1189 eV photon energy).
			The composite image in \textbf{d} shows the sum of detector frames taken before laser excitation ($t < \qty{0}{ps}$, left) and frames taken after laser excitation ($\qty{35}{ps} < t < \qty{685}{ps}$, right).
			For data analysis, the 2D detector frames are azimuthally integrated as shown in \textbf{c}. The 1\textsuperscript{st} and 3\textsuperscript{rd} order maxima of the scattering intensity are indicated in both panels.
		}
		\label{fig:setup}
	\end{figure*}
	
	The instrument for the time-resolved \ac{SAXS} experiment is sketched in Fig.~\ref{fig:setup}.
	We focus laser pulses from a thin-disk amplifier~\cite{Tummler2009} onto a spinning tungsten target to generate broadband, \qty{9}{ps} pulses across the entire soft-X-ray range~\cite{Mantouvalou2015}.
	Two exchangeable \acp{RZP}~\cite{Brzhezinskaya2013}, tailored for the Fe~L$_{2,3}$ and Gd~M$_{4,5}$-edges at around \qty{715}{eV} and \qty{1200}{eV}, respectively, capture, vertically disperse, and horizontally focus less than \qty{10}{eV} bandwidth of the soft-X-rays through a magnetic sample.
	However, resonant magnetic scattering originates exclusively from a narrower band (FWHM approximately \SI{2}{eV}) near the absorption edges~\cite{Kortright2013}. 
	
	A hybrid-pixel area detector~\cite{Bergamaschi2018} acquires the resulting scattering pattern in 2D.
	With its fast readout rate, synchronised to the 100-Hz repetition rate of the \ac{PXS} driver laser, the detector captures and processes an individual frame for each soft-X-ray pulse.
	We enable the detection of single photon events by minimising the detector's noise floor 
	and by applying a dedicated photon-finding algorithm~\cite{Cartier2014}.
	We mitigate source fluctuations in the time-resolved experiments by modulating the repetition rate of the photoexcitation to \qty{50}{Hz} with a mechanical chopper.
	We can then normalise the SAXS signal of the photoexcited sample to frames recorded without laser excitation and substantially increase their \ac{SNR}.
	Both the capability of single-photon counting and shot-to-shot normalisation are essential to enable the first time-resolved \ac{SAXS} experiments around \qty{1}{keV} on a laboratory scale.
	The delay, $t$, between the IR and soft-X-ray pulses is adjusted by a mechanical delay line and $t = \qty{0}{ps}$ corresponds to the temporal overlap of both pulses. 
	Details of the experimental setup, sample preparation, and data evaluation are given in the Methods section.
	
	%%%%%%%%%%%%%%%%%%%%%%%%%%%%%%%%%%%
	%%%%%%%%%%%%%%%%%%%%%%%%%%%%%%%%%%%
	%%%%%%%%%%%%%%%%%%%%%%%%%%%%%%%%%%%

	\subsection{Full-scale time-resolved resonant SAXS experiment}
	
	Fig.~\ref{fig:setup}d shows two-dimensional \ac{SAXS} data recorded at the Gd~M$_5$ absorption edge (\qty{1189}{eV} photon energy), demonstrating the ability of our laboratory instrumentation to detect resonant scattering from magnetic domains with high dynamic range. 
	The data is presented as a composite image with the sample's ground state on the left and the transient laser-excited state on the right and is accumulated from \num{242500} individual frames ($\approx \qty{40}{min}$ acquisition time) for each of the two scattering patterns.
	In our scattering geometry under normal incidence, the scattering contrast originates from the \ac{XMCD}, which is proportional to out-of-plane magnetisation in the Gd layers~\cite{Kortright2013}. 
	As a result, the square root of the integrated scattering intensity is proportional to the local magnetisation of the domains, $M(t)$.
	As expected, the domain pattern gives rise to a \ac{SAXS} intensity distribution of circular shape, indicating the absence of an in-plane anisotropy in the domain pattern, in agreement with the \ac{MFM} data (Fig.~\ref{fig:setup}b).
	The weak azimuthal anisotropy ($q_x$ vs. $q_y$) observed in the 2D scattering pattern in Fig.~\ref{fig:setup}d arises solely from the focusing characteristics of the \ac{RZP} X-ray optics. 
	These instrumental effects can be independently characterised and corrected. 
	While azimuthal information can offer valuable insight into the reorganisation of magnetic domain patterns, it is not relevant for the isotropic FeGd multilayer studied here. 
	Therefore, we perform a full azimuthal integration of the 2D scattering patterns to extract the transient \ac{SAXS} intensity, $I(q, t)$.
	We plot the integrated intensity in Fig.~\ref{fig:setup}c, showing that our instrument is even sensitive enough to resolve the first and third diffraction orders simultaneously.
	Even diffraction orders are suppressed due to the symmetry of the domain pattern, as the characteristic domain width is equal for up- and down-magnetised domains in remanence.
	Typically, we count between $\qtyrange[]{1e3}{2e3}{ph/s}$ in the 1\textsuperscript{st}-order \ac{SAXS} peak before laser excitation at the Fe~L$_3$ and Gd~M$_5$ absorption edges.
	These intensities agree well with the estimated 1\textsuperscript{st}-order magnetic scattering efficiency on the order of \num{e-3} for the investigated FeGd sample, and the different absorptions in the non-magnetic parts of the sample, as well as variations in detection efficiency at both soft-X-ray energies. 
	Due to the two orders of magnitude lower intensity of the 3\textsuperscript{rd}-order scattering, we focus on the 1\textsuperscript{st}-order peak, whose position, $\qfirst$, reflects the average periodicity of the magnetic domains~\cite{Bagschik2016}.
	
	\begin{figure*}[htbp!]
		\centering
		\includegraphics[width=1.0\linewidth]{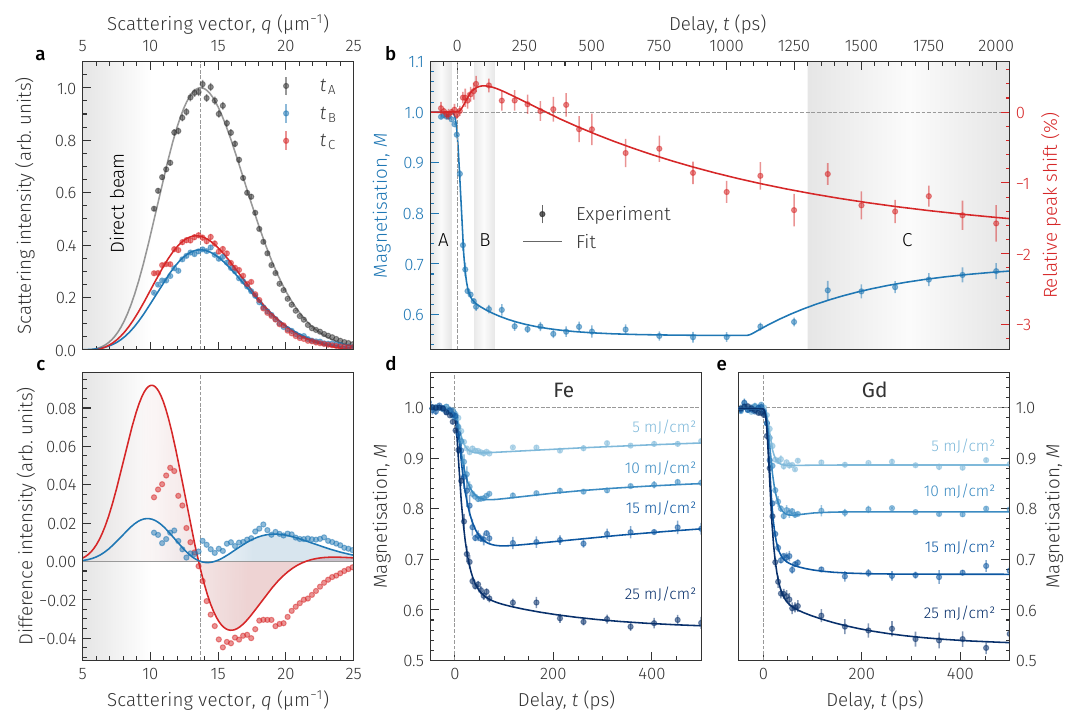}
		\caption{\textbf{Results of the time-resolved resonant SAXS experiment following the laser-driven dynamics of magnetic domains in the FeGd multilayer.}
			\textbf{a},~The $q$-dependent, transient \ac{SAXS} intensity, $I(q, t)$, at the Fe L$_3$ absorption edge at an incident fluence of \qty{25}{mJ\per cm\squared}. 
			Data are averaged for three selected delay intervals A, B, and C as indicated by shaded areas in panel~\textbf{b} and then fitted by a $\Gamma$-distribution (lines). % to access the relevant peak parameters. 
			The grey dashed line indicates the peak position before photoexcitation.
			\textbf{b},~Extracted magnetisation and relative peak shift in dependence of pump--probe delay.
			Data are fitted by the sum of three exponential functions to extract the relevant time scales (lines).
			\textbf{c},~Differences of the \ac{SAXS} intensities, $I(q, t)$, for better visualisation of the peak shift. 
			The data and fits from panel \textbf{a} are normalised by their respective peak intensities before subtracting the data before photoexcitation (A) from the data at positive peak shift (B, blue) and negative peak shift (C, red).
			Markers refer to differences of the experimental data, and lines refer to the differences of the fits.
			Data points were additionally smoothed by a running average.
			\textbf{d,e},~Delay scans of the transient magnetisation, $M(t)$, probed at the two absorption edges Fe~L$_3$ (\textbf{d}) and Gd~M$_5$ (\textbf{e}) for a series of excitation fluences.
			The data is fitted by a double-exponential function to extract the relevant time scales.
			All error bars correspond to the statistical standard error of the mean.
		}
		\label{fig:data}
	\end{figure*}
	
	Upon photoexciting the sample with IR laser pulses, ultrafast demagnetisation reduces the magnetisation and hence the \ac{SAXS} intensity~\cite{Pfau2012, Fan2022}.
	This effect is already directly visible in the composite detector image in Fig.~\ref{fig:setup}d.
	We further fit the azimuthally integrated transient SAXS intensity, $I(q, t)$, obtained at the Fe L$_3$ edge for an incident IR fluence of \qty{25}{mJ/cm\squared} with a model for maze domain patterns~\cite{Bagschik2016} and obtain excellent agreement of the experimental data with this model, see Fig.~\ref{fig:data}a.
	In addition to the reduced scattering intensity after photoexcitation, we observe a peak shift, which is a clear indication of transient spatial modifications of the domain structure.
	
	For a direct and generic determination of the parameters describing the azimuthally integrated SAXS intensity, we numerically extract the transient magnetisation, $M(t)$, from the square root of the integrated intensity $\sqrt{\int{I(q, t)\,\text{d}q}}$, and the peak position, $\qfirst$, from the centre of mass of $I(q, t)$, see Methods section.
	In Fig.~\ref{fig:data}b, we show the time evolution of magnetisation and the peak shift, both normalised to their values before photoexcitation.
	After fast demagnetisation, the sample remains almost unchanged at a magnetisation of $M=0.56$ up to a delay of approximately \qty{1}{ns} before remagnetisation sets in. 
	We have phenomenologically modelled the data with the sum of three exponentials with different time constants, indicating a very fast and a slower demagnetisation process after excitation and the following remagnetisation.  
	We attribute the slower demagnetisation process and its saturation to the fact that the sample thickness is twice as large as the optical absorption length at the laser wavelength of \qty{1030}{nm}.
	Accordingly, direct photoexcitation does not affect the lower part of the multilayer sample, away from the surface on which the laser is incident.
	The low electronic heat conductivity in amorphous FeGd-based materials compared to other crystalline metals~\cite{Hopkins2012} also inhibits efficient short-range transport of hot electrons, which essentially limits the temperature equilibration to the speed of phononic heat transport.
	One-dimensional heat-diffusion simulations~\cite{Schick2021} along the depth of the FeGd multilayer film indeed show a quasi-instantaneous temperature rise and subsequent cooling of the upper half of the film; see Extended Data Fig.~\ref{fig:Sfig1} and Methods.
	In contrast, the lower half of the film reaches a much lower temperature maximum at around $t = \qty{1}{ns}$, indicating that it remains close to its initial magnetisation state.
	At the same time, the average temperature of the film stays nearly constant after photoexcitation for several hundreds of picoseconds before efficient cooling to the substrate becomes effective.
	
	\subsection{Extracting nanoscale structural information}
	
	Information on transient spatial rearrangements of the domain structure is encoded in the position and shape of the diffraction peak.
	A shift of the scattering peak has been observed on ultrafast time scales in Co-based systems and has been controversially discussed in the community~\cite{Pfau2012, Bergeard2015, Zusin2022, Hagstrom2022, Jangid2023}. 
	In contrast to these results, the relative peak shift in our experiment on FeGd exhibits a distinctly different temporal behaviour compared to the magnetisation. 
	First of all, we observe a so far unseen reversal of the transient peak shift from an initial shift to larger scattering vectors superseded by a shift to smaller vectors, i.e. into the opposite direction, as highlighted by the regions B and C in Fig.~\ref{fig:data}b, respectively. 
	While the initial shift to increased $q$ follows the fast demagnetisation time scale, the subsequent peak shift to lower $q$ lasts over the full delay range recorded and is unaffected by the magnetisation recovery.
	
	Comparing the peak shapes of the \ac{SAXS} signal, $I(q, t)$, provides further insight, as shown in Fig.~\ref{fig:data}a.
	We implement this comparison by calculating the difference of the intensity-normalised peaks before and after laser excitation separately for regions B and C in Fig.~\ref{fig:data}c. 
	The diffraction peak at late times (C, red) shows the typical signature of a peak shift with an increase in intensity on one side of the peak and a decrease on the other side compared to the initial peak shape before excitation. 
	We, therefore, attribute the shift mainly to a transiently modified lateral domain periodicity, most likely caused by the warming of the entire multilayer on long time scales. 
	
	The change of the peak at early 100-ps time scales (B, blue) is much more subtle, though. 
	The difference plot shows an increase in intensity on both sides of the peak but with different amplitudes.
	This collective broadening and shifting of the \ac{SAXS} peak indicate spatially strongly inhomogeneous dynamics of the magnetic domain pattern, which is in line with the distinctively different thermal time scales for the upper and lower half of the FeGd multilayer and the corresponding thermal gradient within the depth of the sample.
	We assume that the shift then originates from the transiently reduced effective thickness of the film, promoting local domain-wall rearrangements towards a smaller periodicity~\cite{hehn1996, Montoya2017}. 
	
	Although we can only speculate on the micromagnetic origins of the reversal behaviour of the peak shift, our results demonstrate the wealth of information that can be extracted from the $q$ and time-resolved data accessible by our laboratory setup.
	While in laser-driven non-equilibrium the complex interplay of magnetic interactions can lead to reorganisation and formation of lateral magnetic order on femto- to nanosecond time scales~\cite{Iacocca2019, Buttner2021, Bergeard2015, Hagstrom2022, Pfau2012, Zusin2022, Jangid2023}, our observations highlight the importance of thermal gradients and phononic heat transport in the picosecond regime.
	
	\subsection{Multidimensional measurements}
	
	Studying complex multi-component magnetic materials requires multi-dimensional measurements including variations of applied field, temperature, excitation parameters, and spectral information, e.g., to dissect different sublattices. 
	The high SNR of our data, based on the single-photon counting capabilities and shot-to-shot normalisation, enables highly photon-efficient experiments. 
	We demonstrate this high efficiency and stability by a full set of delay scans obtained at the Fe~L$_3$ and Gd~M$_5$ edges, respectively, with varying excitation fluence, see Fig.~\ref{fig:data}d,e.
	An example of a magnetic hysteresis scan is presented as Extended Data Fig.~\ref{fig:Sfig2}.
	The resonant probing at the two absorption edges allows us to distinguish the magnetic moments in both sublattices~\cite{Radu2011}.
	Typical scans require on average only \qty{100}{s} of integration time per data point ($10^4$ X-ray pulses), which is equivalent to the total acquisition time due to the fast read-out of the detector.
	As an example,  our delay scans typically consist of 40 data points without re-optimising the \ac{PXS} settings, which results in a total duration of maximum \qty{60}{min} per scan.
	The acquisition of the eight scans presented can be completed in a single day, thereby ensuring the comparability of the results.
	
	The transient magnetisation, $M$, in the delay scans in Fig.~\ref{fig:data}d,e is determined from the integrated \ac{SAXS} intensity in the same way as described for panel b. 
	Qualitatively, both sublattices show the same transient change in magnetisation.
	All traces feature a fast initial drop due to ultrafast demagnetisation after photoexcitation, where the demagnetisation rate almost linearly depends on the excitation fluence in the range covered in our experiment.
	However, on average, we observe a \qty{4.5}{\percent} stronger initial demagnetisation at \qty{50}{ps} of the Gd-sublattice magnetic moments as compared to the Fe moments.
	Such differences in the magnetisation dynamics of the two sublattices in TM-RE multilayers or alloys are well known in the literature~\cite{Koopmans2010, Wietstruk2011, Radu2011}.
	These differences are typically even more pronounced in the time constants of the demagnetisation in the first few picoseconds.
	However, our limited soft-X-ray pulse duration of \qty{9}{ps} does not allow us to resolve such differences.
	Still, we detect that remagnetisation occurs faster for the Fe sublattice compared to Gd, which points to a weaker magnetic coupling via interlayer exchange across the approximately \qty{0.5}{nm} thin individual layers as compared to earlier work~\cite{Bartelt2007}.
	
	\section{Discussion and Perspective}
	
	The results on the laser-excited dynamics of magnetic domains in FeGd prove the remarkable capabilities of our laboratory-based instrument for resonant \ac{SAXS} experiments.
	Our instrument is specifically designed for time-resolved studies with a temporal resolution of \qty{9}{ps} defined by the duration of the X-ray pulses. 
	Although the contrast of a magnetic SAXS experiment in transmission at normal incidence is based on a circular dichroism (\ac{XMCD}), the method does not suffer from the random polarisation of the \ac{PXS} emission, as both circular polarisation states contribute equally to the scattering pattern and overlap incoherently.
	The source and beamline provide the required lateral coherence length ($>\SI{500}{nm}$) and photon flux (\SIrange[]{e6}{e7}{ph/s/eV}) at the sample position, to record diffuse scattering of unprecedented intensity and data quality for a laser-driven setup working in the soft-X-ray range between \SIrange[]{500}{1500}{eV} photon energy.
	The detection and normalisation schemes introduced in our instrument allow reasonable measurement times to perform even multidimensional and systematic studies in a tailored laboratory environment.
	For the future, we see great potential for an increase of photon flux of up to two orders of magnitude, as current thin-disk lasers and area detectors~\cite{Tremsin2021} can operate already today at \qty{10}{kHz}, compared to the repetition rate of \qty{100}{Hz} in our experiment.
	
	But what is the actual benefit of our instrument compared to existing setups?
	So far, laser-pump/soft-X-ray-probe \ac{SAXS} experiments were only possible at dedicated beamlines and instruments at \ac{XFEL} facilities.
	In recent years, the technique has led to notable discoveries in the field of photo-induced ultrafast magnetisation dynamics, such as spin currents excited during ultrafast demagnetisation as well as all-optical switching~\cite{Pfau2012,Graves2013} and the picosecond formation, rearrangement, and topological transformation of nanoscale magnetic order~\cite{Iacocca2019,Buttner2021,Zusin2022,Hagstrom2022}.
	However, the flexibility and accessibility of these instruments are largely limited, preventing time-consuming systematic studies within the large phase space of most complex materials.
	Although the temporal resolution of the \ac{PXS} does not capture the fastest sub-picosecond dynamics of magnetic order, it is ideally suited for resolving the intrinsic time scales associated with processes involving spatial rearrangements in non-equilibrium states of technologically relevant meso-scale material systems, spanning the pico- to nanosecond regime.
	Similar studies are possible at \acp{XFEL}, but they will compete fiercely for beamtime with those experiments that are not feasible without the femtosecond pulse durations of \acp{XFEL}, which are even approaching the attosecond regime~\cite{Hartmann2018}.
	
	In contrast, synchrotron-radiation facilities provide X-ray pulses with a typical duration of only \qty{100}{ps}.
	In addition, their high MHz repetition rates render laser-pump/soft-X-ray-probe experiments very challenging because of the requirement to match the pump and probe repetition rates.
	To handle the heat load on solid materials from the laser excitation, approaches at these sources require the reduction of the X-ray repetition rate to typical kHz frequencies via mechanical choppers or dedicated fill patterns of the storage ring~\cite{Collet2003}.
	Laser-slicing facilities on the other hand provide \qty{100}{fs} X-ray pulses at kHz repetition rates~\cite{Holldack2014} with a photon flux of \qtyrange{e6}{e7}{ph\per s \per eV} at the sample being comparable to our \ac{PXS}-based instrumentation.
	However, due to the requirement of nanosecond electronic gating, no 2D detectors have been implemented at soft-X-ray slicing facilities yet, which is inevitable for efficiently covering a larger part of reciprocal space in time-resolved \ac{SAXS} experiments.
	
	Laboratory \ac{SAXS} experiments in the \ac{XUV} regime driven by \ac{HHG} have been able to access nanometre-scale dynamics of magnetic domains with sub-\qty{100}{fs} temporal resolution~\cite{Vodungbo2012, Fan2022}.
	However, the strong absorption of \ac{XUV} light puts narrow boundaries on the maximum sample thickness.
	For higher-lying resonances, which are accessible by our \ac{PXS}, these thickness constraints are much more relaxed and allow for the study of a broader range of fundamentally and technologically relevant sample systems.
	In the soft-X-ray range, the accessible reciprocal space is also about ten times larger and comes along with a much clearer spectral separation of absorption edges, resulting in an unambiguous assignment of signals to their respective elements.
	Although \ac{HHG} sources already reach the soft-X-ray regime, their photon flux is still insufficient for time-resolved \ac{SAXS} experiments at magnetic TM L-edges and above~\cite{Buades2018, Kroh2018, Schoenlein2019, Pupeikis2020, VanMorbeck-Bock2021}.
	
	Due to the short wavelengths of $\lambda \approx \qty{1}{nm}$ combined with the high selectivity of soft X-rays to charge, spin, and orbital degrees of freedom, we are confident that our approach for time-resolved soft-X-ray \ac{SAXS} experiments in a laboratory environment enables a wide range of experimental studies on the ultrafast dynamics of emergent textures in quantum materials.
	Since the scattering contrast in \ac{SAXS} experiments generally does not depend on the source polarisation, as discussed here for the \ac{XMCD} contrast in magnetic domains, also other types of nanoscale domains of, e.g., different morphologies or oxidation states~\cite{johnson2023}, and even ferroelectric domains exhibiting \ac{XLD}~\cite{butcher2025}, are accessible by the randomly polarised \ac{PXS}.
	Our setup's high flexibility allows for easy adaptation to these different material classes, e.g., by controlling external parameters such as fields and temperature.
	Dedicated schemes for photoexcitation employing a synchronised, femtosecond laser and subsequent frequency-conversion stages are promising next steps for accessing dynamics up to the millisecond regime, introducing electronic delays, and for resonantly exciting quasi-particles, such as excitons, phonons, or magnons, respectively.
	
	\newpage
	
	\section{Methods}
	\acresetall
	\subsection{Laser system}
	Our laser system is an in-house developed thin-disk laser system designed to produce high-energy laser pulses at \qty{100}{Hz} repetition rate with exceptional stability. 
	In a first stage, pulses with \qty{100}{fs} pulse duration are picked at \qty{100}{Hz} repetition rate from a Yb:KGW oscillator and then grating-stretched in time to nanosecond duration and amplified to about \qty{1}{mJ} in a first Yb:YAG regenerative amplifier.
	Our Yb:YAG-based thin-disk laser system further amplifies the pulses to an energy in excess of ${\qty{200}{\milli\joule}\,\pm\,\qty{0.2}{\percent\,RMS}}$~\cite{Tummler2009}.
	The pulses are then compressed to a pulse duration of \qty{1.8}{ps} with an efficiency of $\approx\qty{80}{\percent}$.
	From this laser at a wavelength of \qty{1030}{nm}, up to \qty{2}{mJ} are out-coupled for optical excitation of the sample.
	The remaining majority of $\approx\qty{160}{mJ}$ is used to drive the following \ac{PXS}.
	
	\subsection{Soft-X-ray source and beamline}
	For the \ac{PXS}, we focus the laser to a spot size of \qty{15}{\micro\metre} (FWHM) onto a spinning tungsten cylinder.
	Here, every single laser pulse creates a plasma, emitting broadband soft-X-ray pulses of \qty{9}{ps} duration into the full solid angle~\cite{Borchert2023b}.
	The soft X-rays from this source are captured, dispersed, and focused through the magnetic sample by one of two \acp{RZP}, designed for photon energies around \qty{715}{eV} and \qty{1200}{eV}, respectively, as the single optical element onto an area detector.
	In our experiment, the photon energy bandwidth is defined by the sample aperture.
	Optionally, a motorised exit slit can be used to further reduce the bandwidth upstream of the sample.
	An in-vacuum electromagnet provides an applied magnetic field of up to $B=\qty{1.5}{T}$.
	A closed-cycle helium cryostat (not used in this experiment) is installed to control the sample temperature down to \qty{30}{K}.
	
	In this experiment, the sample is photoexcited by 1030-nm IR pulses of \qty{1.8}{ps} duration split off from the \ac{PXS} driving laser.
	The IR pulses are guided over a mechanical delay stage, enabling pump-probe delays of up to \qty{2}{ns}.
	Subsequently, they are focused onto the samples down to a spot size of \qtyproduct{800 x 800}{\micro\metre} with a maximum pulse energy of up to \qty{1}{mJ}.
	An optomechanical chopper synchronised to the thin-disk laser blocks every second pump pulse to capture frames without photoexcitation for normalisation.
	
	\subsection{MÖNCH detector \& data reduction}
	
	The soft-X-ray pulses scattered off the sample are recorded by the hybrid pixel detector MÖNCH (Jungfrau Micropixel with enhanced pOsition rEsolution usiNg CHarge integration) developed at the Paul Scherrer Institute, Switzerland~\cite{Bergamaschi2018,Ramilli2017}. 
	The detector has an active area of \qtyproduct{10 x 10}{\milli\metre} with \qtyproduct{400 x 400}{\pixel} and is located at a distance of \qty{607}{mm} from the sample. 
	The entire chip is coated with a 450-nm-thick Al layer to prevent IR light from reaching the photosensitive region.
	The readout of the entire chip takes \qty{250}{\micro\second} and the data can be transferred at a maximum frame rate of \qty{3}{kHz}.
	The high frame rate allows us to retrieve the scattering image for every soft-X-ray pulse emitted by the \ac{PXS} at its \qty{100}{Hz} repetition rate. 
	We achieve temporal synchronisation of the detector acquisition window and the arrival of the soft-X-ray pulses on the detector with only small jitter ($<\qty{50}{\nano\second}$), which allows us to reduce the detector exposure time to its minimum of \qty{100}{\nano\second}. 
	At this exposure time, we achieve a minimum readout noise of \qty{33}{electrons} (RMS)~\cite{Ramilli2017}.
	In contrast to previous characterisation in Ref.~\cite{Ramilli2017}, the noise of our prototype significantly increases by a factor of more than two when increasing the exposure time to \qty{1}{ms}, which would make single-photon detection in the soft-X-ray range impossible.
	For example, in our case, a 707-eV photon (Fe L edge) creates a signal of \qty{193}{electrons}, and a 1189-eV photon (Gd M edge) a signal of \qty{325}{electrons}.
	The quantum efficiency of the detector was determined at the PTB in Berlin, Germany, to \qty{66}{\percent} and \qty{90}{\percent} at the Fe and Gd edge, respectively.
	We find that a photon is usually split at a ratio of $\approx 60/40$ between two pixels horizontally or vertically, which we account for by using a photon-finding algorithm based on a continuous comparison of pixel values and the sum of \qtyproduct{2 x 2}{\pixel} and \qtyproduct{3 x 3}{\pixel} sub-regions with the RMS-weighted threshold values~\cite{Cartier2014}. 
	Prior to this processing, we first remove the background (pedestal) from the raw data and apply a common-mode correction to account for a charge redistribution between illuminated and non-illuminated pixels.
	The resulting scattering patterns are then sorted and summed up according to the state of the optomechanical chopper in the pump beam path.
	The two-dimensional intensity patterns are further azimuthally integrated via the Python \textsc{pyFAI}~\cite{Kieffer2013} package.
	We can fit the $q$-dependent SAXS peak, $I(q)$, by an, e.g., $\Gamma$-distribution-based fit model~\cite{Bagschik2016}
	\begin{equation}
		f(q, A, \theta, k, c) = A\cdot{\frac {1}{\Gamma (k)\theta ^{k}}} q^{(k-1)}\exp{\left(-{\frac {q}{\theta }}\right)} + c,
	\end{equation}
	where the scattered intensity is directly represented by the amplitude $A$ and the mean value, representing the average domain size, is given by  $q_\text{mean} = \theta k$.
	
	For a direct and generic determination of the parameters describing the azimuthally integrated \ac{SAXS} intensity, we numerically extract the local magnetisation of the domains as square root of the scattered \emph{intensity} $M \propto \sqrt{\int{I(q, t)\,\text{d} q}}$~\cite{Kortright2013} as the scattering \emph{contrast} is proportional to the magnitude of the magnetisation in the domains with alternating magnetisation direction.
	The peak position, $\qfirst$, is determined by the centre of mass of $I(q, t)$.
	The analytical $\Gamma$-based model and the numerical methods both result in qualitatively similar results for the transient changes of the peak area and position.
	In addition to single-photon events dominating the diffuse scattering, the high frame rate of the detector also enables a sufficiently large dynamic range to simultaneously capture the intensity of the direct, forward-scattered X-ray beam in the centre of the scattering pattern, see Fig.~\ref{fig:setup}d.
	
	The transient data sets of the magnetization $M(t)$ and the peak position $\qfirst(t)$ are fitted by a sum of exponential functions to account for the various different time scales observed.
	The most general form of this function is given by
	\begin{equation}
		f(t) = A_0 + G(t) \ast \left\{\sum_i  \Theta(t-t^0_i) C_i \left( 1 - e^{-(t-t^0_i) / \gamma_i} \right) \right\}
	\end{equation}
	where $A_0$ is a static offset, the temporal resolution is included by convolution by a Gaussian function $G(t)$, the Heaviside step function $\Theta(t-t^0_i)$ describes the delay-dependent onset of the different dynamics at $t^0_i$, and the exponential behaviour is described by a scaling parameter $C_i$ and a time constant $\gamma_i$.
	
	\subsection{Sample system}
	
	Our sample system is an [Fe(\qty{0.4}{nm})$\vert$Gd(\qty{0.5}{nm})]$_{116}$ multilayer, where the alternating Fe and Gd layers couple antiferromagnetically to each other.
	The multilayer exhibits magnetic stripe domains with an average size of $\approx \qty{232}{nm}$ at room temperature, forming a maze pattern with alternating up and down out-of-plane magnetisation; see \ac{MFM} image in Fig.~\ref{fig:setup}b.
	The magnetic heterostructure was grown by electron-beam physical vapour deposition on an X-ray-transparent \qty{20}{nm}-thick SiN membrane of \qtyproduct{0.5 x 0.5}{\milli\meter} aperture, seeded with \qty{3}{nm} and capped with \qty{2}{nm} of Ta.
	The backside is coated with a \qty{300}{nm}-thick layer of Al for improved heat dissipation to reduce static heating effects.
	Out-of-plane hysteresis loops probing the magneto-optical Kerr effect (MOKE) and SAXS intensity at the Fe L$_3$ edge clearly evidence the transition of a domain state at remanence to a fully saturated ferromagnetic state at external fields $B > \qty{75}{mT}$, see Fig.~\ref{fig:Sfig2}.
	
	\subsection{One-dimensional heat diffusion simulations}
	
	To estimate the time and length scales of energy transport within the FeGd sample after laser excitation, we carried out one-dimensional heat diffusion simulations using the Python \textsc{udkm1Dsim} package~\cite{Schick2021}.
	For the accessible time scales ($t > \qty{10}{ps}$), we describe electrons and phonons with a single temperature, $T$, assuming thermal equilibrium between the two subsystems.
	We solve the one-dimensional heat-diffusion equation
	\begin{equation}
		c \rho \frac{\partial T}{\partial t} = \frac{\partial}{\partial z} \left( k \frac{\partial T}{\partial z} \right)  + S(z, t)
	\end{equation}
	with isolating boundary conditions.
	Here, $c$ denote the material-dependent specific heat capacities of each layer and $k$ the respective thermal conductivities.
	For simplification, $c$ and $k$ are assumed to be independent from $T$.
	The optical excitation is taken into account by $S(z, t)$, describing the pump laser energy absorbed by the sample.
	To this end, the depth-dependent differential absorption profile of the \qty{1030}{nm} pump pulses is first calculated using a multilayer formalism within the \textsc{udkm1Dsim} framework, taking interlayer reflection and interference effects into account, resulting from the material-dependent complex refractive index change at each interface of the heterostructure.
	The spatial absorption profile of the pump pulses is further multiplied by a Gaussian distribution in time with an experimentally determined duration of \qty{1.8}{ps} (FWHM).
	The simulation code and all material-specific parameters are provided within a dedicated repository.
	
	\backmatter
	
	\section{Acknowledgement}
	We acknowledge funding from the Leibniz Association through the Leibniz Junior Research Group Grant No.\ J134/2022 and from the Deutsche Forschungsgemeinschaft (DFG) through TRR 227, Project No.\ A02.
	
	\section{Competing interests}
	The authors declare no competing interests.
	
	\section{Data availability}\label{sec:data_availability}
	
	The data and simulation code reported in this study are available with identifiers at \url{https://doi.org/10.5281/zenodo.17225735}.
	
	\section{Author contributions}
	B.P. and D.S. conceived the study. M.B. and D.S. developed the soft-X-ray instrument. 
	L.L., B.P., and D.S. characterised the detector and developed the integration into the instrument and the DAQ software.
	L.L., M.B., N.S. performed the measurements with support from K.K., M.S., and D.S.; 
	D.E. and L.L. produced and characterised the sample. 
	N.S. and D.S. performed the heat-transport simulations. L.L., M.B., B.P., and D.S. analysed and interpreted the data with support from M.S.;
	L.L., M.B., B.P., and D.S. wrote the manuscript with input from all authors.
	Supervision of students by S.E. and D.S.
	Funding was acquired by S.E. and D.S.
	
	\bibliography{bibliography_SAXS}
	
	\clearpage
	\acresetall
	
	\begin{appendices}
		\section*{Extended Data}
		
		\renewcommand{\figurename}{Extended Data Fig.}
		
		\renewcommand{\thefigure}{E\arabic{figure}}
		\setcounter{figure}{0}
		
		\begin{figure}[htbp!]
			\centering
			\includegraphics[width=1\linewidth]{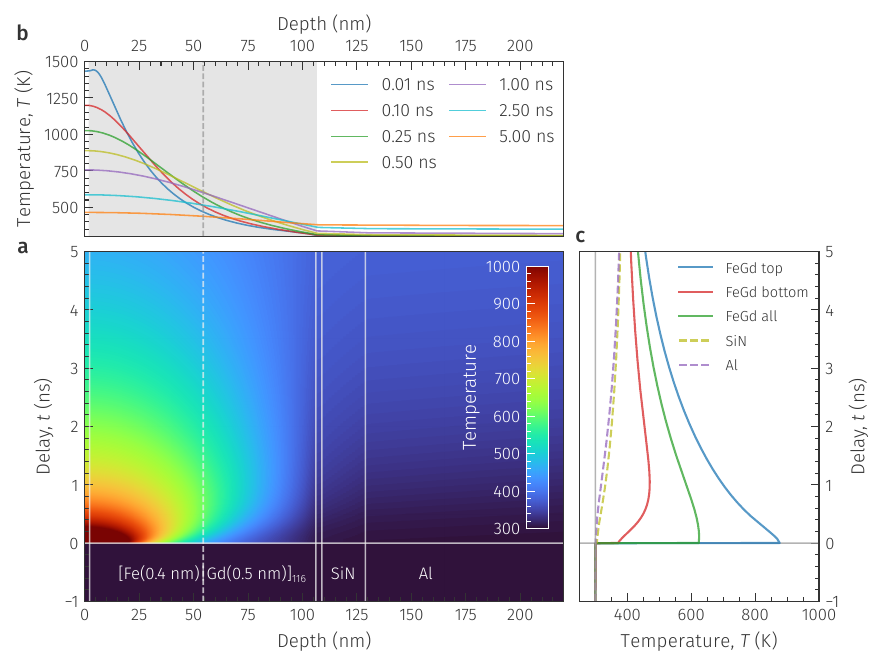}
			\caption{\textbf{One-dimensional heat diffusion simulation after laser excitation of the investigated FeGd heterostructure.}
				Details of the simulations are described in the Method section.
				Values for the electronic and phononic thermal conductivities of FeGd are taken from Ref.~\cite{Hopkins2012}.
				\textbf{a}, Time-dependent temperature distribution within the entire sample (cap layer, magnetic multilayer, SiN substrate, Al heat sink).
				The vertical solid grey lines indicate the top and bottom of the actual magnetic multilayer, and the vertical dashed grey line its centre dividing the layer in a ``top'' and ``bottom'' part (see panel c).
				\textbf{b}, Line-outs of the spatial temperature distribution for different pump-probe delays $t$ as indicated.
				\textbf{c}, Average temperature of different regions of the sample heterostructure as function of the pump-probe delay $t$.
			}
			\label{fig:Sfig1}
		\end{figure}
		
		\begin{figure}[htbp!]
			\centering
			\includegraphics[width=0.7\linewidth]{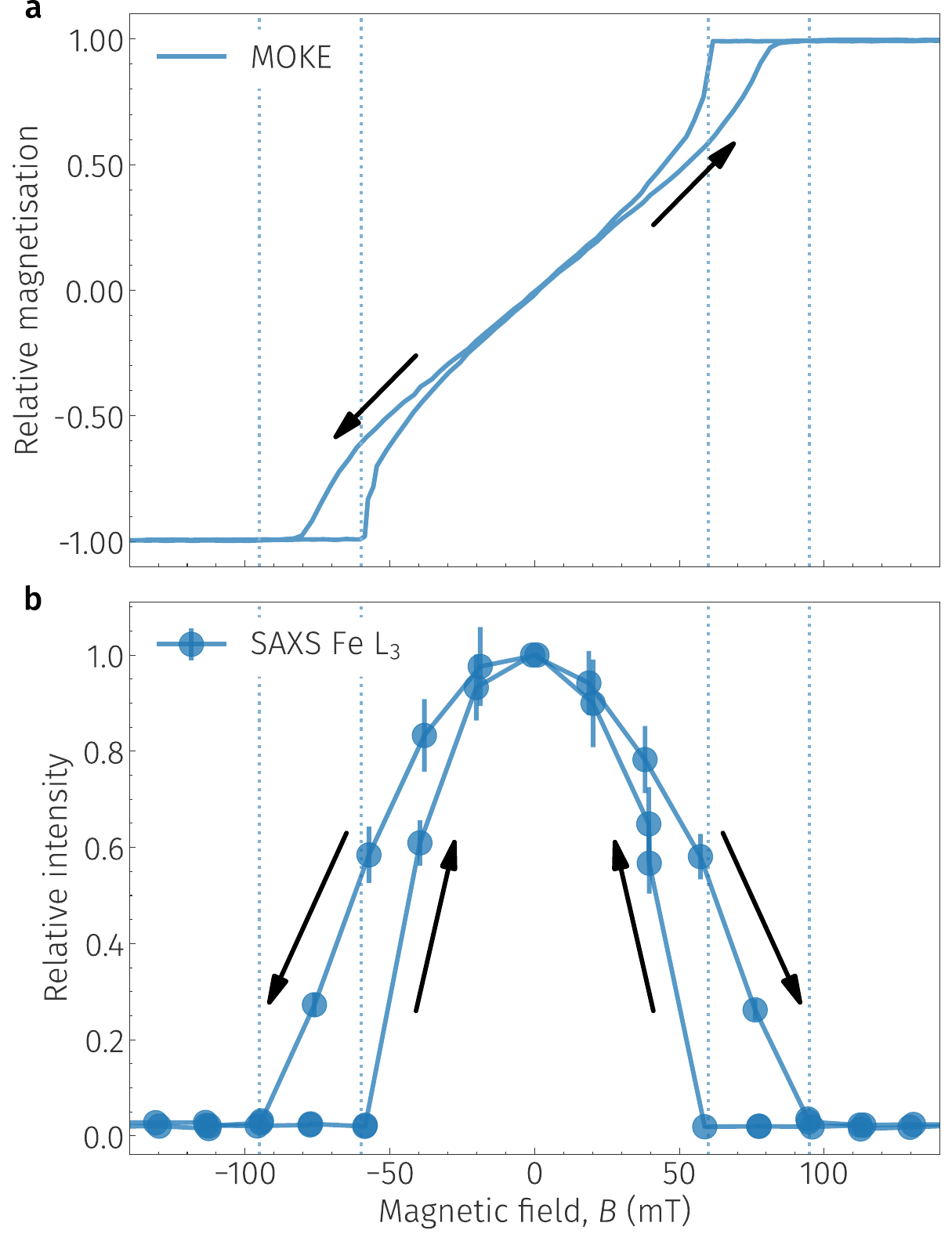}
			\caption{\textbf{Hysteresis scan of the investigated FeGd heterostructure.}
				The black arrows indicate the scan direction of the applied magnetic field, $B$.
				\textbf{a}, Magneto-optical Kerr-effect (MOKE) measurement of the net out-of-plane magnetisation.
				In remanence, no net magnetisation is detectable as the contributions from oppositely magnetised domains cancel out.
				At larger fields ($B > \qty{75}{mT}$) all domains align parallel into ferromagnetic saturation.
				\textbf{b}, The SAXS peak intensity at the Fe L$_3$ absorption edge exhibits maximum contrast approximately in remanence.
				The scattering intensity vanishes at high fields when the domain structure disappears in ferromagnetic saturation.
			}
			\label{fig:Sfig2}
		\end{figure}
		
	\end{appendices}
	
\end{document}